\documentclass[prd,preprint,eqsecnum,nofootinbib,amsmath,tightenlines]{revtex4}

\def\M{{\cal M}}

\def\p{{\bf p}}

\def\alphas{\alpha_{\rm s}}
\def\LQCD{\Lambda_{\rm QCD}}
\def\GeV{{\; \rm GeV}}

\def\ca{C_{\rm A}}
\def\cf{C_{\rm F}}
\def\cs{C_{\rm S}}

\def\tf{t_{\rm F}}
\def\ts{t_{\rm S}}

\def\nf{N_{\rm f}}
\def\ns{N_{\rm s}}

\def\nc{N_{\rm c}}
\def\ng{N_{\rm g}}

\def\Eq#1{Eq.~(\ref{#1})}

\def\st{\begin{equation}}
\def\stp{\end{equation}}
\def\bg{\begin{eqnarray}}
\def\nd{\end{eqnarray}}
\def\nn{\nonumber}
\def\hs{$\;\;$}
\def\hso{$\phantom{0}\;\;$}
\def\hsoo{$\phantom{00}\;\;$}

\def\lsim{\mbox{~{\protect\raisebox{0.4ex}{$<$}}\hspace{-1.1em}
	{\protect\raisebox{-0.6ex}{$\sim$}}~}}

\def\slashchar#1{\setbox0=\hbox{$#1$}           
   \dimen0=\wd0                                 
   \setbox1=\hbox{/} \dimen1=\wd1               
   \ifdim\dimen0>\dimen1                        
      \rlap{\hbox to \dimen0{\hfil/\hfil}}      
      #1                                        
   \else                                        
      \rlap{\hbox to \dimen1{\hfil$#1$\hfil}}   
      /                                         
   \fi}                                         %
\def\nott#1{\slashchar{#1}}
\def\Llv{L_{{\rm L}\!\!\! /\,}}
\def\Plv{P_{{\rm L}\!\!\! /\,}}

\advance\parskip 1.9pt
\advance\voffset -0.2in

\begin{document}


\title{Limits on Lorentz violation from the highest energy cosmic rays}

\author{Olivier Gagnon and Guy D.\ Moore}
\affiliation
    {%
	Department of Physics,
	McGill University, 3600 University St.,
	Montr\'{e}al QC H3A 2T8, Canada
    }%

\date{\today}

\begin{abstract}
We place several new limits on Lorentz violating effects, which can modify
particles' dispersion relations, by considering the highest energy
cosmic rays observed.  Since these are hadrons, this involves
considering the partonic content of such cosmic rays.  We get a number
of bounds on differences in maximum propagation speeds, which are
typically bounded at the $10^{-21}$ level, and on momentum dependent
dispersion corrections of the form $v = 1 \pm p^2/\Lambda^2$, which
typically bound $\Lambda > 10^{21}$ GeV, well above the Planck scale.
For (CPT violating) dispersion correction of the form $v = 1 +
p/\Lambda$, the bounds are up to 15 orders of magnitude beyond the
Planck scale.
\end{abstract}



\maketitle


\section {Introduction}

It is generally believed that Lorentz invariance is an exact symmetry of
nature.  This belief is supported by extremely precise experimental
tests, and by strong and well motivated theoretical prejudice.  Indeed,
exact Lorentz invariance is used as one of the cornerstones on which
relativistic quantum field theory is built, leading to the extremely
successful Standard Model of Particle Physics.  Together with the
equivalence principle, local Lorentz invariance is one of the
assumptions underpinning General Relativity.  Alternatively, we could
say that General Relativity is the gauge theory of general coordinate
invariance, and has exact local Lorentz invariance as a consequence.
In either case, it is intimately related to both our best theories of
particle physics and of gravity.

While most physicists believe that Lorentz symmetry is an exact symmetry
of nature, one of our jobs as physicists is to put all of our
assumptions to the most vigorous possible tests.  Therefore,
we should attempt to place the tightest possible
limits on the violation of Lorentz symmetry.
A general framework for considering Lorentz violation has been developed
by Kosteleck\'y and collaborators \cite{Kostelecky}.
The best laboratory experimental limits come
from clock comparison experiments, for instance the recent experiments
of Cane {\it et.~al.\ } \cite{Cane}, see also \cite{Clock_expt}.
These experiments are highly constraining, particularly for low energy
effects and those involving CPT symmetry violation.
(In quantum field theory, CPT symmetry is a
consequence of Lorentz symmetry, together with causality, locality,
analyticity, and unitarity, and its violation implies that Lorentz
symmetry is also violated \cite{Greenberg}.)

However, it is consistent and reasonable to
believe that CPT could be an exact symmetry without Lorentz invariance.
It is also desirable to obtain limits involving rather high energies.
In fact, it turns out high energies introduce another way of gaining
high precision, in a manner suggested by Coleman and Glashow
\cite{ColemanGlashow,ColemanGlashow2}.

The argument goes as follows.  If Lorentz symmetry is not
exact, then different particles can have different limiting velocities,
and potentially more complicated corrections to their dispersion
relations.  This opens up the possibility of Cherenkov radiation; a
particle with a larger limiting velocity is energetically allowed to
radiate particles with lower limiting velocities, even though such
radiation by a particle at rest is not kinematically allowed.  For
instance, suppose the limiting velocity of an electron is $c_e >
c_\gamma$, where $c_\gamma$ is the limiting velocity of a photon.  Then
the energy of an electron of momentum $p \gg m c_e$ is
\st
E_{e,p} = \sqrt{ p^2 c^2_e + m_e^2 c_e^4} \simeq p c_e +
	\frac{m_e^2 c_e^3}{2p} \, ,
\stp
while the energy of a photon of momentum $k$ is $k c_\gamma$.  The
process $e \to e\gamma$ is kinematically allowed if the final state
energies sum to the initial state energy and the sum of the magnitudes
of the final state momenta exceeds the magnitude of initial state
momentum (so a nonvanishing opening angle is allowed).  Denoting the
final photon momentum as $k$, defining
$\epsilon \equiv (c_e-c_\gamma)/c_\gamma$, and taking
$m_e^2 c_e^2/p^2 \sim \epsilon \ll 1$, this condition becomes
\st
\mbox{Cherenkov condition:} \quad
k \epsilon > \frac{m^2 k c_e^2}{2p(p-k)}\mbox{ for some $k$}
	\, , \quad \mbox{or} \quad
\epsilon > \frac{m^2 c_e^2}{2p^2} \, ,
\stp
which is the same as the condition that the electron is traveling faster
than the speed of light $c_\gamma$.  This is ``traditional'' Cherenkov
radiation.

The highest energy particles observed or inferred are not those
accelerated in laboratory, but those which occur in cosmic rays or other
astrophysical settings.  The Cherenkov process is so
efficient that an electron which exceeds the threshold for Cherenkov
radiation would lose energy down to the threshold on a time scale which
is very short compared to any astrophysical time scale.  Therefore,
electrons of higher energy would never exist astrophysically.
One can then place a limit on $\epsilon$ of
\st
\epsilon < \frac{m_e^2 c^2}{2p^2} \, ,
\stp
with $p$ the momentum of the most energetic electron observed or
inferred in an astrophysical setting.  (Since
the constraint on $\epsilon$ will be so tight, there is no need to
specify which $c$ appears on the right hand side.)

Similarly, if the photon's limiting velocity is larger, then the process
$\gamma \to e^+ e^-$ becomes kinematically allowed for photons of
momentum $k$ satisfying
\st
\mbox{Pair production condition:}\quad
k c_\gamma > k c_e + \frac{2 m_e^2 c^3}{k} \, ,
\stp
where we already used that the process is kinematically most favored
when the electron and positron share the momentum evenly.  This leads to
a bound,
\st
- \epsilon < \frac{2 m_e^2 c^2}{k^2} \, ,
\stp
with $k$ the momentum of the most energetic photon observed.
The process involved here is not ``traditional'' Cherenkov radiation,
but a sort of ``generalized'' Cherenkov process.

Observations of cosmic ray photons, and of photons inferred to arise
from synchrotron emission off electrons in the Crab nebula, give the
constraints \cite{Crab_limits}
\st
-2\times 10^{-16} < \epsilon < 5\times 10^{-20} \, ,
\stp
a two-sided bound on Lorentz
violation (though not as strong as those from clock experiments).
The authors of \cite{Crab_limits} also show that such
Cherenkov limits place bounds on dispersion corrections involving
additional powers of the particle momentum, of form $E = pc +
p^2/\Lambda$, which limit $\Lambda$ to be at or above the Planck scale
(a constraint much stronger than can be obtained from existing
laboratory experiments).

One would expect that the strongest limits of this sort would arise from
the cosmic rays with the largest energy.  The most energetic cosmic rays
observed carry energy in excess of $10^{11}$ GeV
and are believed to arise from hadronic primaries \cite{Cosmic_Rays}.
The highest energy cosmic ray observed had an energy of about $3\times
10^{11}$ GeV \cite{Fly_eye}, and was probably a hadron \cite{fly_hadron}.
The fact that such high energy particles can arrive from astronomical
distances gives the constraints \cite{ColemanGlashow,MooreNelson},
\begin{eqnarray}
\frac{c_p - c_\gamma}{c_\gamma} & < & 10^{-23} \, , \\
\frac{c_p - c_{\rm grav}}{c_p} & < & 10^{-15} \, ,
\end{eqnarray}
where $c_{\rm grav}$ is the propagation speed of gravitation.
(The constraint involving gravitation is weaker because gravitational
Cherenkov radiation is inefficient; the velocity difference must be
$10^{-15}$ before gravi-Cherenkov radiation becomes appriciable on an
astrophysical time scale.)
Unfortunately, these bounds are both one-sided; $c_p$ could be
substantially slower than $c_\gamma$ or $c_{\rm grav}$ without violating
either bound.

The proton is a composite particle.  At the energy relevant to the
current problem, $10^{11}$ GeV, the proton (as viewed in our rest frame)
can be considered as a loose
bag of several partons, mostly quarks and gluons, each carrying a
small fraction of the proton's total energy.  The purpose of this paper
is to show that, by taking into account the partonic structure of
hadrons, we can use the arrival of high energy hadrons from astronomical
distances to place a number of severe limits on Lorentz violating
physics.  For the reader's convenience, we will present the main results
here.

In all of our results, we will assume that the highest energy extensive
air showers arise when single hadrons strike the upper atmosphere.  (It
is possible that heavier nuclei are actually responsible; this would
weaken some of our bounds.)  For
some results, we will make an additional assumption; in a ``bottom-up''
generation mechanism for the high energy cosmic rays.  That is, we will
sometimes assume that the highest energy cosmic rays got their energy by
being accelerated by ordinary
electromagnetic fields over astronomical distances, rather than, for
instance, arising as the decay products of some long lived and
ultra-heavy relic particles.  In this case, the hadron must have been
electrically charged during the acceleration process.  Therefore,
some charged hadron must be stable, at least on the time scale of hours,
at the energy of the highest energy cosmic rays observed.
In what follows, we will label results which depend on this
assumption with an asterisk $*$.

Differences in propagation speeds arise from dimension 4 Lorentz
violating operators.  We will parameterize these by the propagation speed
difference.  Choosing to renormalize such operators at the scale
$10^{11}$ GeV (see below), we find the following limits:
\begin{eqnarray}
-1.6\times 10^{-23} & < \frac{c_1 - c_{1/2}}{c_1} < &
	1.4 \times 10^{-23} \, , \\
* \quad -1.8\times 10^{-21} & < \frac{c_q - c_g}{c_g} <
	& 1.2\times 10^{-21} \, , \quad * \\
* \quad -2\times 10^{-21} & < \frac{(c_q {+} c_g)/2 - c_\gamma}{c_\gamma} < &
	5 \times 10^{-24} \, , \\
* \quad -1\times 10^{-20} & < \frac{(c_q{+}c_g)/2 - c_e}{c_e} < &
	5 \times 10^{-24} \, .
\end{eqnarray}
In the first line, $c_1$ and $c_{1/2}$ are the propagation speeds of
fundamental gauge bosons and fundamental fermions, if propagation speeds
are common to particles of a given spin but differ for particles of
different spins.
In the second line, $c_q$ is the propagation speed for quarks assuming
all quarks share a maximum velocity, and $c_g$ is for gluons.  In the remaining
lines, $c_\gamma$ and $c_e$ are the propagation speeds for photons and
electrons (the physical particles) respectively; the limits on electrons
also apply for neutrinos.

Note that all of these limits are two sided, and many do not require an
assumption about the nature of cosmic ray acceleration.  These
constitute severe constraints on Lorentz violation.  In deriving these
limits we have assumed that the $3\times 10^{11}$ GeV cosmic ray was a
hadron and that its energy was correctly measured; the strength of the
bound involves the square of the cosmic ray's energy, so if the highest
energy cosmic ray hadron energy were only $1 \times 10^{11}$ GeV, the
results are an order of magnitude weaker.

Dimension 5 operators induce spin dependent $O(p^2/\Lambda)$
corrections to particle dispersion relations \cite{Pospelov}, with
opposite sign for particle and anti-particle (and for opposite
polarization states of gauge fields).
There is an independent such parameter for each Standard Model field.
Note that these operators violate CPT, and so can be avoided by
insisting on CPT symmetry.

Unless the coefficient $k$ of these corrections, $E = p \pm k
p^2$, are at least 20 times larger for QCD degrees of freedom than for
photons and leptons, we can obtain the following constraints on $k$ for
photons and leptons:
\begin{eqnarray}
|k_\gamma| & < & \frac{1}{10^{34}\GeV } \, , \nn \\
|k_{\nu_e}| , |k_{\nu_\mu}| , |k_{\nu_\tau}| & < &
	\frac{1}{10^{34}\GeV } \, , \nn \\
|k_{e_R}|, |k_{\mu_R}| & < & \frac{1}{10^{34}\GeV } \, , \nn \\
|k_{\tau_R}| & < & \frac{1}{5\times 10^{32}\GeV } \, .
\end{eqnarray}
These limits are about 15 orders of magnitude stronger than the Planck
scale, and do not depend on assumptions about the origins of cosmic
rays.  This means that any theory which induces dimension 5, CPT
violating operators must generate them at a highly trans-Planckian
scale, or must introduce them with a very small coefficient $\lsim
10^{-15}$.  However, for certain peculiar corners of the
parameter space of dimension 5 operators, we cannot place limits.  For
instance, if the right-handed up quark receives the largest dispersion
correction and it is in the range
$E = p -k_{u_R} p^2$, with $k>1/(10^{32}\GeV)$; and if dispersion
corrections of all non-QCD particles are at least 30 times smaller, then
the $\Delta^{++}$ may be the most stable propagating particle at the
energies of the highest energy cosmic rays, and no limits can be placed.

We are able to place super-Planck scale limits (assuming the Lorentz
violating physics introduces the operators with $O(1)$ coefficients)
even on dimension 6 operators.  Such operators induce
CPT-even dispersion corrections of form $E = p + K p^3$.  The limits we
obtain on the coefficient $K$ are,
\bg
\frac{-1}{(4\times 10^{22} \GeV)^2} & < K_1 < &
	\frac{1}{(6\times 10^{21} \GeV)^2} \, , \nn \\
\frac{-1}{(4\times 10^{22} \GeV)^2} & < K_{1/2} < &
	\frac{1}{(1.6\times 10^{22} \GeV)^2} \, , \nn \\
*\quad \frac{-1}{(6\times 10^{21} \GeV)^2} & < K_q < &
	\frac{1}{(1.6\times 10^{22} \GeV)^2} \, , \nn \\
*\quad \frac{-1}{(2\times 10^{21} \GeV)^2} & < K_g < &
	\frac{1}{(6\times 10^{21} \GeV)^2} \, , \nn \\
\frac{-1}{(4\times 10^{22} \GeV)^2} & < K_\gamma < &
	\frac{1}{(3\times 10^{21} \GeV)^2} \quad
	\mbox{or} \quad 25 |K_q| \quad * \, , \nn \\
\frac{-1}{(4\times 10^{22} \GeV)^2} & < K_{\rm lept} < &
	\frac{1}{(1\times 10^{21} \GeV)^2} \quad
	\mbox{or} \quad 500 |K_q| \quad * \, .
\nd
Here $K_1$ and $K_{1/2}$ are the dispersion corrections of spin-1 and
spin-1/2 particles if these are taken to be common to all particles of
the same spin; $K_q$ is the
dispersion correction of quarks if it is taken to be flavor independent,
and $K_{\rm lept}$ is the largest of the dispersion corrections of any light
lepton ($e$, $\mu$, $\nu_e$, $\nu_\mu$, or $\nu_\tau$).  In the last
two lines, the bound is whichever limit is weaker.

We believe that these constraints will make it next to impossible to
develop a theory in which Lorentz symmetry is entirely absent, since
they imply a scale of violation at least 2 orders of magnitude higher
than the Planck scale.  They are rather less constraining if Lorentz
symmetry is spontaneously broken in some hidden sector.
Note that the scale appearing in the bound arises as $p^2/m_p$, with $p$
the momentum of the highest energy cosmic ray; the number quoted should be
scaled as the second power of that energy, if $3\times 10^{11}$ GeV
does not hold up as the energy of the highest energy cosmic ray.  Some
of these bounds, such as the lower bounds on $K_1$ and $K_\gamma$, apply
even if the cosmic rays are nuclei rather than individual hadrons.

The remainder of the paper is organized as follows.  In Section
\ref{sec:ops}, we review and discuss what Lorentz violating operators
look like and how they generate corrections to dispersion relations.  In
Section \ref{sec:twist}, we show that the corrections to a hadron's
dispersion relation involve the expectation value of the operator in a
state given by a highly boosted hadron.  This in turn is related to a
twist two operator which has a simple expression in terms of the parton
distribution functions (PDFs) of the hadron, evaluated at a scale set by the
parton's energy {\em in our frame}.
In Section \ref{sec:pdfs}, we present our results for integral moments
of parton distribution functions renormalized to the scale $10^{11}$
GeV.
Section \ref{sec:results} shows how the
resulting PDFs can lead to specific decay processes which generate the
constraints we have just presented.
We end with a short conclusion.  Finally, details on PDF evolution
in the Standard Model are postponed to an appendix.

\section{Lorentz violating operators}
\label{sec:ops}

It is fair to assume that any Lorentz violating effects are small, in
which case we can treat Lorentz violation as a small perturbation on
Lorentz invariant ordinary physics.  In the language of field theory,
Lorentz violation can be described by the introduction of Lorentz
violating operators into the (Standard Model) Lagrangian.

Fairly general discussions of Lorentz violating operators have been
presented by Colladay and Kosteleck\'y in a general context, and by
Coleman and Glashow in the context of the Standard Model
\cite{ColemanGlashow}.  The most convenient
way to introduce such operators is to introduce one or more preferred
frame 4-vectors.  One then constructs Lagrangian terms as usual, except
that Lorentz indices can be contracted against preferred frame 4-vectors
as well as against each other.

Consider for instance electrodynamics with a fermionic field $\psi$ and
a gauge boson $A^\mu$, with a single timelike 4-vector $u^\mu$, $u^\mu
u_\mu = -1$, defining
a preferred frame%
\footnote{We use [ -- + + + ] metric convention.}.
Besides the usual Lagrangian terms,
\st
m \bar\psi \psi + \bar\psi \gamma^\mu D_\mu \psi
+ \frac{1}{4e^2} F_{\mu \nu} F^{\mu \nu}
\, ,
\stp
we can write the new dimension 4 terms,
\st
(c_\psi^{-1}-1) u_\mu u_\nu \bar\psi \gamma^\mu D^\nu \psi
+\frac{c_\gamma^{-2}-1}{4e^2} u_\mu u^\nu F^{\mu \alpha} F_{\nu \alpha} \, ,
\stp
which modify the free velocities of propagation, in the frame where
$u^\mu$ is a unit time vector, as indicated.  We can eliminate one or
the other term by performing a rescaling of the coordinate system,
stretching the ``time'' coordinate (the coordinate which is time in the
frame where $u^\mu$ points in the time direction), so that one particle
velocity again becomes 1.  However, the ratio of velocities for the
different fields does not change under such a coordinate redefinition,
so if this ratio is not 1, then Lorentz violation cannot be eliminated
by any coordinate transformation.%
\footnote
    {%
    It is worth remarking that other considerations, such as causality
    and locality, may limit what terms can appear in such a Lagrangian
    \cite{Kostelecky2}.  For instance, Kosteleck\'y and Lehnert argue that
    the term we have written, with $v<1$, cannot be canonically
    quantized in all frames, while the $v>1$ case may have causality
    violation issues \cite{Kostelecky2}.%
    }

It is also possible to write Lorentz violating operators of both lower
and higher dimension.  For instance, a dimension 3 term,
\st
u_\mu \bar\psi \gamma^\mu \psi \, ,
\stp
splits the energies of the particle and antiparticle, a CPT violating
effect.  At dimension 6, a term of form
$-K u_\mu u_\alpha u_\beta u_\gamma \bar\psi \gamma^\mu \partial^\alpha
\partial^\beta \partial^\gamma \psi$ modifies the dispersion relation, so
that $E-KE^3$, rather than $E$, appears (in the frame where $u^\mu$ is
purely temporal).  At high energy and treating the
perturbation to be small, the dispersion relation in the rest frame is
then approximately $E \simeq p + K p^3$, which perturbs the propagation
velocity by $v = 1+3 K p^2$ (setting $c=1$).

The coefficients of Lorentz violating operators will run with scale,
and will generically mix.  Rather than investigate this in detail, we
will constrain their values at the scale set by the physics which is
providing the constraint, which in our case will be $\sim 10^{11}$ GeV.

The absolutely most general Lorentz violating Lagrangian has enough
terms to make the analysis very complicated.  We will make a number of
simplifying assumptions:
\begin{enumerate}
\item
There is a unique timelike 4-vector $u^\mu$ appearing in Lorentz
violating Lagrangian terms, and its preferred frame is not too different
from the frame of the microwave background (the boost factor is at most
$O(1)$);

\item
For the most part, we will only consider terms even in $u^{\mu}$.
Terms odd in $u^\mu$ violate CPT, giving opposite corrections to a
particle and its CP conjugate.

\item
Lorentz violating effects will be family diagonal.  We will also present
certain constraints assuming they are family blind, or even that they
are the same for all gauge bosons and for all fermions.

\end{enumerate}
The last assumption is for convenience, and because we don't know of any
strong motivation for theories where it is violated.  The other
assumptions do not seriously restrict our conclusions, because the cases
where they do not hold are typically more strongly constrained.
For instance, dropping CPT symmetry allows dimension 3 operators, on
which there are very tight laboratory constraints \cite{Cane}.
Nevertheless, we will consider what happens when it is violated--though
our ignorance of spin dependent parton distribution functions will make
it hard to make as complete an analysis as we make for the CPT
symmetric case.  Regarding the first assumption, we make two
comments.  Constraints on spacelike 4-vectors are about as strong as
those on timelike 4-vectors because we have observed ultra-high energy
cosmic rays arriving over a wide swath of the sky.  For the case where
$u^\mu$ is purely temporal (or purely spatial) only in a frame of
reference at a large boost with respect to our frame, some of the cosmic
rays we observe have much higher energies in that frame, and would
therefore Cherenkov radiate even more strongly.  Therefore such
$u^\mu$ are more tightly constrained than the $u^\mu$ which is at rest
in the microwave background frame.

To summarize, Lorentz violating effects behave, at dimension 4, like
species dependent speeds of light, and at dimension 6, like shifts
in the propagation speed of form $\delta v \sim \p^2/\Lambda^2$.
Because it is convenient and intuitively clear, we will generally
parameterize the Lorentz violating operators in terms of their dispersion
relation corrections in the remainder of the paper.

\section{Hadron dispersion and the twist expansion}
\label{sec:twist}

How does a Lorentz violating perturbation to the Lagrangian $\Llv$
modify a particle's dispersion relation?

First, we must derive from $\Llv$ the perturbation to the 4-momentum
operator $\Plv^\mu$.  The perturbation to the particle's 4-momentum is then
given by
\st
\delta p^\mu = \langle \psi,\p | \, \Plv^\mu \, | \psi,\p \rangle
	\, ,
\label{eq:deltap}
\stp
where $\psi$ is the particle type and $\p$ is the particle's momentum.
The velocity is $v = dp^0/d|\p|$, so it is modified, in the
ultra-relativistic and small Lorentz violation limit, by
$\delta v \simeq d(\delta p^0 - \delta |\p|)/d |\p|$.  Since we can
shift $\Llv$ by a Lorentz conserving piece, we can generally make
$\delta |\p|=0$, in which case the correction is just $d\delta p^0/d|\p|$.
Evaluating this for a free particle is straightforward; we find the part
of $\Llv$ which looks like a kinetic term for this particle, and replace
derivatives with the momentum $p^\mu$.  The free particle relation between the
Lagrangian terms and the modification of particle dispersion relations
then holds.

But what if the particle in question is a hadron?
Naively, we should view the hadron as a sum of partons with different
momentum fractions $x$ of the hadron's momentum, and expect $\delta
p^\mu$ to be the sum of the contribution expected from each
parton.  That is, suppose that the dispersion relation for parton type
$f$ is
\st
\delta p^0[f,\p] = \langle f,\p |\, \Plv^0 \,| f,\p \rangle
	= (c_f-1) |\p| + K_f |\p|^3 \, .
\stp
Write the parton distribution function for a hadron $h$, such that the
probability to find a parton $f$ with momentum fraction between $x$ and
$x{+}dx$ when using a probe of virtuality $q^2$ is given by
$f_h(x,q^2) dx$.  We should then expect the correction to the hadron's
energy to be,
\st
\delta p^0[h,\p] = \langle h,\p |\, \Plv^0 \,| h,\p \rangle
= \sum_f \int_0^1 dx \Big[ (c_f - 1) x |\p| +
	K_f x^3 |\p|^3 \Big] f_h(x,|\p|^2) \, .
\label{eq:expectation}
\stp
This basically turns out to be the case.

For our case, $\Plv$
is simplest in the rest frame of the 4-vector $u^\mu$, and in this
frame, $|\p| \gg \LQCD$, showing that the expectation value of the
Lorentz violating operator is determined by hard physics.
Alternatively, we could say that
$u^\mu p_\mu$ provides a hard scale which makes a partonic description
applicable.  When evaluating an operator such as $\Plv$ in the wave
function of a boosted hadron, as in \Eq{eq:deltap}, one should perform
an operator product expansion on $\Plv$ in terms of low twist operators.
The dominant contribution
will arise from the twist-two part of the operator, with higher twists
suppressed by powers of $\LQCD / u^\mu p_\mu \sim 10^{-11}$.

Consider a twist two, dimension $D$ operator ${\cal O}_D$.  Denote its
expectation value in an unphysical state with parton type $f$ carrying
momentum $p$ to be
\st
\langle f,\p | \, {\cal O}_D \, | f,\p \rangle
	\equiv \langle {\cal O}_D \rangle_f \, .
\stp
Then the {\em definition} of the parton distribution functions (PDFs)
is, that the expectation value of ${\cal O}_D$ in a hadron is given at
leading order by
\st
\langle h,\p | \, {\cal O}_D \, | h,\p \rangle =
	\sum_f \langle {\cal O}_D \rangle_f
	\int_0^1 dx \, x^{D-3} f_h(x) + O(\alphas/2\pi) \, ,
\stp
where $f_h(x)$ is the PDF for parton $f$ in the
hadron $h$.  This is valid up to higher order effects, which are
minimized (order $\alphas/2\pi$ without logarithmic enhancement)
if we take the operator ${\cal O}$ to be renormalized at the relevant
hard scale, which for us is about $10^{11}$ GeV.%
\footnote
    {%
    The Lorentz violating operators are presumably induced at a scale of
    order the Planck scale and must then be RG evolved from that scale
    to the scale $10^{11}$ GeV.  Such evolution has been studied for the
    case of QED in \cite{Kostelecky3}.  The evolution generically leads
    to $O(\alpha \log(\Lambda/\mu))$ mixing between operators, making it
    unlikely that one operator should be smaller than another of the
    same dimension by more than about a factor of 10.  One also expects
    high dimension operators of dimension $D$ to induce lower dimension
    ones of dimension $d$ with coefficient $\sim \alpha \Lambda^{D-d}$.
    This is suggests that dimension 4 operators will have
    coefficients $O(\alpha)$ \cite{Collins}, which as we will see is
    grossly in conflict with observation.
    }  
In other words,
the expectation values of twist-two operators are given by the Mellin
moments of the PDFs--or to be more accurate,
at leading order the PDFs are defined as the
inverse Mellin transforms of the twist-two moments of the hadron.

Therefore, the correct statement is that the value of an operator in a
hadron is the sum over partonic species of the integral over $x$ of the
PDF times the expectation of the operator for the given parton.
In other words, our naive expectation in \Eq{eq:expectation} is correct,
up to $O(\alphas/2\pi)$ corrections.  Since $\alpha_s \sim 0.03$ at the
scale in question, we can safely ignore such corrections.  (The
corrections could be found by making a loop-level treatment of the
operator product expansion of the Lorentz violating operator in the
twist expansion, and a loop level evaluation of the twist-two operator
in terms of the PDFs of the hadron.)

Therefore, the effect of dimension 4 Lorentz violating operators is to
give a hadron a limiting velocity of propagation of
\st
c_{h} = \sum_f c_f \int_0^1 dx \, x f_h(x) \, ,
\stp
with the sum running over all partonic species.
(Note that $\sum_f \int_0^1 dx\, x f_h(x) = 1$ to ensure momentum
conservation.)  The limiting velocity of a hadron is the momentum
weighted average of the naive limiting velocities of its partonic
constituents.  Similarly, if parton type $f$ naively gives a dispersion
relation of $E = p + K_f p^3$ due to dimension 6 Lorentz violating
operators, then the energy of a hadron will be given by
\st
E = p + p^3 \sum_f K_f \int_0^1 dx \, x^3 f_h(x) \, .
\stp
Since most of the momentum of the hadron is carried at relatively small
$x$, this means that the correction to the hadron dispersion, due to
dimension 6 operators, will be much smaller than it would be were the
hadron not composite.

\section{Standard Model Partonic Content of Hadrons}
\label{sec:pdfs}

The evolution kernel for parton distribution functions, within QCD, is
known to next-to-leading order (NLO) \cite{NLO_PDF}, and the evolution
equations for lower order integral moments are known to NNLO
\cite{NNLO_PDF,Vermaseren}.  Global fits of the PDFs of
the proton have been performed using data at many energy scales and
values of $x$ \cite{CTEQ,MRST,GRV}; those of the neutron are given, to
suitable accuracy, by replacing $u\leftrightarrow d$.
None of the available PDF libraries
go to $10^{11}$ GeV, and none include the evolution of electroweak
degrees of freedom (the PDFs of photons and weak bosons, leptons, and
the Higgs boson within the proton).

Since some of our most interesting results will come
from analyzing the relatively small contamination of electroweak degrees
of freedom in hadrons, it behooves us to extend the analysis in this
way.  Luckily, we are only interested in integral moments of the PDFs,
which actually have much simpler evolution equations with scale.
Namely, the integral moments of form $\int_0^1 x^n f(x) dx$ obey a
closed set of ordinary differential (renormalization group) equations,
which can be evolved along with the renormalization group (RG) equations
for the couplings, once 
initial values for the integral moments are determined from the global
fits.  The appendix explains why the integral moments satisfy ordinary
differential equations (a point which is well known at least to parts of
the community), and extends the usual treatment to include the LO
effects of all Standard
Model interactions, including electroweak interactions with scalars and
Yukawa interactions.

Since we are not generally interested in tracking how much of the
structure is in particles and how much in ant-particles, we will only
keep track of the sum of the PDFs, $q(x)+\bar{q}(x)$ (the singlet
distribution), which we will nevertheless label $q(x)$.  We will also
not track spin dependent PDFs.  The spin dependence and
particle-antiparticle difference are both needed to place detailed
limits on dimension 5 CPT violating operators, but the spin dependent
PDFs are not known with enough accuracy to justify such a treatment
anyway.

We have evaluated the relevant integral moments for the quark and gluon
PDFs in a proton, neutron, and pion at the relatively low scale of 3
GeV, by using a fit from the CTEQ collaboration \cite{CTEQ} for the
$p,n$ and an old MRST collaboration fit \cite{MRST_pion} for the pion.
These were evolved to the scale $10^{11}$ GeV.
In this evolution, gauge field beta functions and QCD contributions to
the DGLAP equations (the equations describing the scale dependence of
the PDFs) were treated at the two loop
level, while the running of
the Yukawa coupling and all other DGLAP evolution was treated at one
loop level.  The Higgs self-coupling was not treated at all, because it
first leads to DGLAP evolution at two loops (and then only in the
uninteresting Higgs scalar sector) and to Yukawa and gauge boson beta
functions at two and three loops respectively.
We also treated all thresholds in a rather naive way; for mass
thresholds, a particle was excluded from the theory until $q^2=m^2$, and
then included as massless.  For the electroweak threshold, taken to
occur at $m_Z=91.19$ GeV, we abruptly switched to treating QED+QCD to
treating the Standard Model, neglecting all particle masses.  At this
threshold, the PDF for the $u$ quark was split evenly between the right
and left handed components (appropriate for a spin averaged sample of
protons), and then the left $u$ type and $d$ type quarks were
mixed with each other, while the photonic PDF content was mixed into the
$B$ and $W$ boson types according to $\cos^2 \Theta_W$ and $\sin^2
\Theta_W$ respectively.  (Mixing left quarks would occur anyway due to
soft $W$ radiation, a process which is logarithmically divergently
efficient.)

This treatment of PDF evolution, as well as the starting PDFs,
naturally introduce errors.  This is particularly true for the rather
poorly known pion PDFs.  Note however that the PDF evolution of the
integral moments in question tends to converge towards a fixed point,
particularly for the strongly interacting degrees of freedom (because of
the large size of $\alphas$ and the consequent rapid evolution of the
PDFs).  This is evidenced by the startling similarity of the proton and
pion PDFs at the ultraviolet scale.  As a consequence, the errors are
reduced.  In addition, in what follows we will be most interested in the
{\em difference} between the PDFs of different particles, and the
errors in these differences--particularly differences arising because of
electroweak interactions--are much smaller.  For these reasons, we will
not make a detailed error analysis of our PDF evolution.  (In any case,
our final errors will be dominated by the error in the energy of the
highest energy cosmic ray observed to date, which may introduce as much
as a factor of 2 error in the numbers in our bounds.)
That said, we should particularly point out that the pion PDFs are not
as well known, in particular the size of the quark sea is uncertain.
Using the extreme large and small sea values given in \cite{MRST_pion},
we find that the gluon fraction in Table \ref{Table:dim4} varies by $\pm
.004$.  Fortunately, we do not find that processes involving pions are
particularly important in bounding Lorentz violating operators.

\begin{table}[t]
\centerline{
\begin{tabular}{|c||c|c|c|c|c|c|c|}\hline
$\quad$ Momentum fraction $\quad$ & $p$ & $n$ & $\pi$ & $\gamma$ &
	$e$ & $D^0$ & $gg$ \\ \hline \hline
$\quad$gluon$\quad$& .454 & .455 & .454 & .019 & .002 & .425 & .491  \\ \hline
1'st gen.\ quark   & .269 & .270 & .266 & .024 & .004 & .122 & .184  \\ \hline
2,3'rd gen.\ quark & .223 & .223 & .227 & .046 & .008 & .391 & .278  \\ \hline
lepton doublet     & .002 & .002 & .002 & .054 & .393 & .003 & .002  \\ \hline
lepton singlet     & .000 & .000 & .000 & .065 & .439 & .001 & .000  \\ \hline
U(1) gauge field   & .009 & .008 & .008 & .627 & .077 & .013 & .007  \\ \hline
SU(2) gauge field  & .040 & .040 & .040 & .159 & .075 & .044 & .035  \\ \hline
Higgs boson        & $\;\;$.003$\;\;$ & $\;\;$.003$\;\;$ & $\;\;.003\;\;$
	& $\;\;$.007$\;\;$ & $\;\;$.001$\;\;$ &
	$\;\;$.002$\;\;$ & $\;\;$.003$\;\;$  \\ \hline
\end{tabular}}
\caption{Dimension 4 content, at $10^{11}$ GeV scale, of $p$, $n$, $\pi$,
$\gamma$, $e$, $D^0$, and glueball ($gg$).
\label{Table:dim4}}
\end{table}

We also evolved the PDFs for the electron, the photon, and crude models of
the $D$
meson and the glueball.  For the electron and the photon, we took the
partonic content to be a pure electron or photon with $x=1$ at the scale
$500$ KeV, and evolved under electromagnetic interactions only to the
scale 500 MeV, and with full interactions from there.  This procedure
will underestimate the QCD content of the photon somewhat; but we find
that neither the photon nor the electron takes on much admixture of
other degrees of freedom, and it is not important in
practice to use the evolved rather than pure structure.

Treating the $D$ and the glueball requires more modeling.
We treat the $D$ meson, at the scale $m_c \sim 1.5$ GeV, as a charm
quark with $x=0.8$, with the remaining $20\%$ of the momentum made up
equally of light quark and gluon.  The light quark and gluon content are
taken to have $\langle x^2 \rangle =0.01$ each, so they contribute quite
little to the high dimension content of the particle.
The structure at higher scales is then obtained by DGLAP running,
starting at the scale $1.5$ GeV.  We considered similar models of the
$J/\psi$ and $B$ mesons, but these turn out to be too heavy to be
important in setting the Lorentz violating bounds.
The glueball was taken to be pure glue at the
scale 0.5 GeV, with $\langle x^2 \rangle = 0.2$, and was evolved from
there.  We think this model of the $D^0$ is reasonable, but the glueball
model is more speculative.

The integral moments we consider can in principle be determined by
lattice gauge theory techniques.  This would be particularly useful for
the glueball, as otherwise we have little guidance in how to treat it.

\begin{table}[t]
\centerline{
\begin{tabular}{|c||r|r|r|r|r|r|r|}\hline
$\quad 10^2 \times \int x^3 f(x) dx \quad$
& $ p \;\quad$ & $ n \;\quad$ & $ \pi \;\quad$ &
	$ \gamma \;\;\;\quad$ & $ e \;\;\;\quad$ & $ D^0 \quad$
	& $ gg  \quad$ \\ \hline \hline
$\quad$gluon$\quad$& .061 \hs & .062 \hs & .126 \hs &   .096 \hs &
	 .004  \hs & .486 \hs & .026 \hs \\ \hline
1'st gen.\ quark   & .427 \hs & .437 \hs & .900 \hs &   .699 \hs &
	 .036  \hs & .092 \hs & .043 \hs \\ \hline
2,3'rd gen.\ quark & .037 \hs & .037 \hs & .076 \hs &  1.35 \hso &
	 .071  \hs &4.14 \hso & .042 \hs \\ \hline
lepton doublet     & .001 \hs & .000 \hs & .001 \hs &  2.33 \hso &
	30.3 \hsoo & .006 \hs & .000 \hs \\ \hline
lepton singlet     & .000 \hs & .000 \hs & .000 \hs &  3.36 \hso &
	38.0 \hsoo & .003 \hs & .000 \hs \\ \hline
U(1) gauge field   & .007 \hs & .004 \hs & .011 \hs & 62.2 \hsoo &
	 1.97 \hso & .083 \hs & .001 \hs \\ \hline
SU(2) gauge field  & .012 \hs & .012 \hs & .025 \hs &  6.68 \hso &
	 1.18 \hso & .113 \hs & .002 \hs \\ \hline
Higgs boson        & .000 \hs & .000 \hs & .001 \hs &   .213 \hs &
	  .008 \hs & .002 \hs & .000 \hs \\ \hline
$\quad$Total$\quad$& \hs .546 \hs & \hs .553 \hs & \hs 1.14 \hso &
	\hs 76.9 \hsoo & \hs 71.6 \hsoo & \hs 4.92 \hso &
	\hs .114 \hs \\ \hline
\end{tabular}}
\caption{Dimension 6 content, at $10^{11}$ GeV scale, of $p$, $n$, $\pi$,
$\gamma$, $e$, $D^0$, and glueball ($gg$); in percent ($\%$) of value
for a particle with the full $x=1$ fraction of the momentum.
\label{Table:dim6}}
\end{table}

We find that all QCD particles are about $45\%$ glue and $50\%$ quarks,
with the remainder made up mostly by gauge bosons.  The glueball is
richer in glue and the $D$ richer in quarks, but not strikingly
so.  The photon and electron are about $80\%$ the original constituent.
As for the $x^2$ weighted structure, in every case the sum over
constituents gives less than 1 (as it must; splitting conserves $\sum x$
but always reduces $\langle x^2 \rangle$.)
However, for the electron and photon, it
is a modest reduction, while for hadrons the reduction is enormous;
weighting with $x^2$, barely anything is left (or alternately, most of
the partonic content has $x<0.1$).  This is particularly so
for the glueball; the root-mean-squared value of $x$ for a
glueball constituent is 
about 0.03.  This is because gluon radiation, which brings down a
particle's $x$, is efficient, especially for gluons themselves.

\section{Derivation of main results}
\label{sec:results}

\subsection{Dimension 4 operators}

Consider first, constraints on dimension 4 Lorentz violating operators.
These look like differences, between particles, in the maximum speed of
propagation.  Since our constraints will always require the Lorentz
violating operators to be subdominant to the Lorentz conserving ones, and
since the particles involved will always be highly relativistic, we will
always work to linear order in velocity differences, and will not
distinguish different particle velocities from each other, and will set
$c=1$, after we have made the leading order expansion.  Similarly,
masses will be taken to correct dispersion relations only at leading
order in $m^2/p^2$.  So a particle with propagation speed $(1+\epsilon)$
and mass $m$ will be taken to have energy
\st
E = \sqrt{(1+\epsilon)^2 p^2 + (1+\epsilon)^4 m^2}
  = p + \epsilon p + \frac{m^2}{2p} \, .
\stp

Consider first a difference in velocity between fermions and gauge
bosons; all fermions have a common propagation velocity at the scale
$10^{11}$ GeV, and all gauge bosons have a different common velocity.
This could happen if the source of Lorentz violation, whatever it were,
distinguished spin but didn't ``care'' about gauge charges or other
internal symmetries.

In this case, it is quite easy to find a two sided bound on the
propagation speed difference.  Call that difference,
$\epsilon \equiv c_1 - c_{1/2}$ (with $c_1$ the gauge boson speed and
$c_{1/2}$ the fermion speed).  Consider first the case $\epsilon < 0$.
In this case, the process $p \to p \gamma$ is kinematically allowed at
suitably high energy.  The proton content is almost exactly half each,
gauge boson and fermion, whereas the photon is about $80\%$ gauge boson
and only $20\%$ fermion (see Table \ref{Table:dim4}).
The energy difference between a proton of
momentum $p$, and a final state with collinear proton of momentum
$p(1-x)$ and photon of momentum $px$, is
\bg
\Delta E & = & (1+0.5 \epsilon)p + \frac{m_p^2}{2p}
	- (1+0.5\epsilon)(1-x)p - \frac{m_p^2}{2p(1-x)}
	-x(1+0.8\epsilon)p
	\nn \\
	& = & p \left[ -0.3 \epsilon x -\frac{x}{1-x}\,\frac{m^2}{2p^2}
	\right] \, .
\nd
When this difference becomes positive, the initial state energy is
larger than the final state energy, meaning that the decay can occur
with a finite opening angle (which will make the final state energy a
bit larger and use up the extra energy).  The optimal value of $x$ for
this to happen is $x\simeq 1$, and the process becomes kinematically
allowed at a threshold,
\st
\epsilon = -\frac{m_p^2}{0.6 p^2}\, .
\stp
When this bound is exceeded by a factor of 2, the proton loses of order
half its energy on a length scale of less than a meter; so the bound on
cosmic rays can be taken essentially to coincide with the kinematic
bound for the process.

Now suppose $\epsilon>0$, so fundamental gauge bosons propagate faster
than fundamental fermions.  In this case, contrary to simpleminded
intuition, the breakup of a proton, $p \to p e^+ e^-$, becomes allowed,
since the electron is about $85\%$ fermion and only $15\%$ gauge boson
(almost exclusively electroweak gauge boson).  The kinematics are almost
the same as for the previous case, since the electron mass is negligibly
small.  Combining the two, the bound on $\epsilon$ is,
\st
-\frac{m_p^2}{0.6 p^2} < ( c_1 - c_{1/2} ) < \frac{m_p^2}{0.7 p^2}
	\, .
\stp
Substituting in $m_p = 0.94$ GeV and $p = 3\times 10^{11}$ GeV gives the
bounds quoted in the introduction.  Note that neither bound relies on
any details of the cosmic ray production mechanism; all we need to know
is that the highest energy cosmic rays are hadrons, and that all hadrons
are essentially even mixes of gauge boson and fermion.

We have not worried, in the above limit, about scalar (Higgs boson)
content, because no low energy particle has much Higgs content.  If for
some reason $c_1$ and $c_{1/2}$ were approximately equal and $c_0$, the
fundamental scalar propagation speed, were more different, we could also
achieve limits, about 200 times weaker, by the same processes, because
the photon contains more, and the electron less, Higgs boson than any
hadron made of light constituents.

These limits can be evaded if we assume that
propagation speed differences are not universal but show more
complicated species dependence.  For instance, suppose that
strongly interacting particles propagated slightly more slowly than
other degrees of freedom.  We can still place a bound, albeit a weaker
one, on the propagation speed difference between quarks and gluons,
$\epsilon' \equiv (c_q - c_g)$, as follows.

Suppose first that gluons propagate more slowly.  Then a gluon-rich
hadron would be energetically more favorable.  A hadron of mass $m_1$
and gluon fraction $g_1$ would split off a daughter hadron of mass $m_2$
and gluon fraction $g_2$ provided the energy difference,
\bg
\Delta E & = & (1-g_1 \epsilon')p + \frac{m_1^2}{2p}
	- (1-g_1\epsilon')(1-x)p - \frac{m_1^2}{2p(1-x)}
	- (1-g_2\epsilon')x p - \frac{m_2^2}{2px}
	\nn \\
& = & x(g_2-g_1) \epsilon' p - \frac{m_1^2}{2p} \, \frac{x}{1-x}
	- \frac{m_2^2}{2px}
\label{eq:both_massive}
\nd
is positive for any $x$.  The most gluon-rich hadron that we have been
able to identify is the glueball, a $0^{++}$ scalar.  There is no
conservation rule to forbid the process $p \to p + gg$ ($gg$ the
glueball).  Taking the glueball mass to be $1.6$ GeV
\cite{Glueball_mass}, we find that the process goes forward, for a
proton with energy equal to the highest energy cosmic ray observed, if
$\epsilon'> 1.2\times 10^{-21}$.

The presence of the process $p \to p + gg$ does not necessarily mean
that cosmic rays could not reach the Earth; they would simply have to be
composed of glueballs.  Even though glueballs are unstable at rest, in
the Lorentz violating scenario under discussion, they would be stabilized
by the Lorentz violating terms when their energy reached that of
the highest energy cosmic rays.  Therefore they could
propagate to the Earth over cosmological distances.  This is provided
they were not destroyed by the microwave background; but in fact this
seems likely, because a glueball should have a small cross-section with
photons.  Therefore, this scenario would probably even evade the GZK cutoff
\cite{GZK}.  However, we can rule it out if we make the additional
assumption that the acceleration mechanism for cosmic rays is by
``ordinary'' electromagnetic fields.  Since the glueball is neutral, it
could never
achieve more energy than the proton (or other charged hadron) which
radiated it, and so the energy limit for the proton is the energy limit
for any hadron, within this assumption and assuming as well that no
other charged hadron would remain stable.  This seems safe, as all
charged hadrons have valence quarks which give them quark content of
order or larger than that of the proton.


In the opposite case, $\epsilon'<0$, where gluons propagate faster, a
quark-rich hadron is energetically favored.  The pion may be, and the
kaon is expected to be, slightly quark-richer than the proton, but we
find that neither is richer by enough to compete with the process
$p \to p D^0$, with the $D^0$ a $c\bar{u}$ or $u\bar{c}$ bound state.
The kinematic limit for the process, from \Eq{eq:both_massive}, allows
it to occur whenever $\epsilon' < -1.8\times 10^{-21}$.  The analogous
process with a $D^{\pm}$ requires 6 MeV more energy in the center of
mass frame, because of $m_{D^+} - m_{D^0}$ and $m_n - m_p$.
We have crudely estimated that, when 6 MeV is available in
the center of mass frame, the center-of-mass lifetime for the process
$p\to pD^0$ is around $10^{-9}$ seconds.  The $p \to p D^0$ process
occurs before there is enough energy to permit $p \to n D^+$, provided
the cosmic ray acceleration takes less than a few hundred seconds (in
our frame), which seems reasonable for most proposed cosmic ray sources.
Therefore the proton will lose energy to a neutral particle, which will
not be accelerated further.  The maximum energy a hadron can obtain is
therefore the energy of a proton at the kinematic limit for this
reaction.  This leads to the limits on $\epsilon'$ given in the
introduction.

We already saw that we can derive a limit when $c_\gamma <
(c_q+c_g)/2$.  We can also set a limit in the opposite case, as
follows.  The proton has a larger photon (electroweak boson) content
than the neutron, and when $c_\gamma > c_q,c_g$, this means the limiting
speed of a proton is faster than that of a neutron, opening up the
process $p \to n e^+ \nu_e$ at high enough energies.
(The reason the proton has more gauge boson than the neutron is, that it
has more up quarks, and the electric charge of the up quark is larger
than that of the down quark; so the quarks in a proton
radiate more electroweak bosons.)
The difference in electroweak boson content, barely discernible in Table
\ref{Table:dim4}, is $0.14\%$ of the proton's momentum.  Neglecting for a
moment the electron and neutrino masses, this is enough to counter the
neutron's larger mass when $\epsilon > .0014 m_p (m_n-m_p)/p^2$.
Naively, this leads to a bound of $2\times 10^{-23}$.

However, this is not the kinematic limit we find, because the timescale
for the process $p \to n e^+ \nu_e$ involves the fifth power of the
available energy.  Together with the time dilation of the reaction, this
leads to a time scale for the reaction, when close to the kinematic
limit, of order $10^{14}$ seconds.  This is shorter than the likely
propagation time of a cosmic ray, but it is much longer than the
expected acceleration time of a cosmic ray, which we will take to be
$10^4$ seconds (of order the time variability of an active galactic
nucleus).  This knocks two orders of magnitude off our limit.  If the
correct time scale is shorter, our limit deteriorates as the $1/5$ power
of the inverse acceleration time.  Note also that this limit relies on
our assumption of ordinary acceleration.  However, the energy involved
is enough that the electron mass is negligible, and corrections from
different leptonic dispersion relations will be too small to matter,
according to constraints we will derive next.

We have already seen how $p \to p e^+ e^-$ gives a strong constraint on
the speed difference of $g+q$ and leptons, when the leptons are slower.
Similar considerations for $p \to p \bar \nu \nu$ prove that this limit
applies both to singlet and doublet neutrinos.  However, the limits on
the case where the leptons propagate more slowly are much weaker; the
analog of the constraint we found for the photon is 15 times weaker,
because the lepton content of a hadron is second order in $\alpha$, so
the difference between the leptonic content of a neutron and proton is
$10^{-4}$.  We actually get a stronger bound by combining the two-sided
constraints on $c_\gamma - (c_q+c_g)/2$, just derived, with the Crab
constraints \cite{Crab_limits} on the electron-photon velocity
difference.  This gives the constraints on the $c_e$, $c_p$ differences
quoted in the introduction.

The case where different quarks have different limiting velocities is
richer, and may be phenomenologically viable; for instance, if the up
quark is slower than all other particles, $p \to \Delta^{++} \pi^-$ may
be allowed, with most of the energy carried by the $\Delta^{++}$.  If,
on the other hand, the slowest propagating quark is one of the heavier
species, say, the charm quark, then $p \to p D^0$ again becomes favored
and the proton loses energy to neutral particles; so comparable bounds
would exist in this case.

\subsection{Dimension 5 operators}

Next, consider the case with nonvanishing dimension 5 operators.  Since
they are odd under CPT, these operators always imply opposite dispersion
corrections for a particle and its $CP$ conjugate, so if the
right-handed (SU(2) singlet) up quark has energy $E_{u_R}=p + k_{u_R}
p^2$, then the left-handed (SU(2) singlet) anti-up has energy
$E_{\bar u_R}=p - k_{u_R} p^2$.  Similarly, opposite polarizations of
gauge bosons have opposite corrections.  The coefficient $k$ is
independent for each distinct Standard Model field; in particular, it is
different for right and left handed up quarks, though it is the same for
left-handed up and down quarks.

Since we consider this possibility rather unlikely, we will not attempt
to be very general in our considerations.  Rather, we will assume
that the order of magnitude of $k$ is the same for each species, or at
least the same within species of a given spin, and try
to set some limits on it.

We do not know the spin dependent PDFs of the proton very accurately.
What we do know is that the $x^2$ moments of PDFs in hadrons are
generally quite small, $<0.05$, because most of a hadron's structure at
the scale $10^{11}$ GeV lies in very small $x$ partons.  Therefore, if
$k$ is comparable for different Standard Model species, then only the
dispersion corrections of leptons and electroweak bosons will be
important.

With this in mind, consider again the process $p \to p \gamma$.  Take
the polarization dependent energy shift for the photon to be $E_\gamma =
p \pm k_\gamma p^2$, with $\pm$ according to whether the photon is left or
right circularly polarized.  The lower energy photon can always be
produced; if the helicity of the proton forbids its production, we may
consider $p \to p \gamma \gamma$, with an extra, low energy photon of
the opposite polarization.  The energy difference for collinear
emission, with a photon momentum fraction of $x$, is
\bg
\label{eq:dim5_2body}
\Delta E & = & p + \frac{m_p^2}{2p} - (1-x)p -\frac{m_p^2}{2p(1-x)}
	-xp +|k_\gamma| x^2 p^2 \, , \nn \\
& = & -\frac{m_p^2}{2p}\: \frac{x}{1-x} +|k_\gamma| x^2 p^2 \, .
\nd
This first vanishes when $|k|=2m_p^2/p^3$, $x=1/2$.  Writing
$k=1/\Lambda$, the scale $\Lambda$ is $p^3/2m_p^2 \simeq 1.5\times
10^{34}$.  The result should be slightly less than this, because the
structure of the photon is not purely an $x=1$ photon; but looking at
Table \ref{Table:dim6}, we see that not much of the structure has
cascaded by this scale, so the correction is quite modest.

We arrive at essentially the same constraint on left-handed leptons by
considering $p \to p \nu \bar{\nu}$; this applies in each generation,
since all neutrinos are light.  For right handed leptons, the
constraints involving electrons are the same, and for muons are weaker
only by a factor of $2/3$, but for $\tau$ leptons they are weaker by a
factor of about 20, due to the mass of the particle to be produced.

In the case where dispersion corrections are more than 30 times larger
for QCD degrees of freedom than for leptons or electroweak bosons, these
constraints do not apply.  New constraints may or may not emerge,
depending on exactly what the corrections are.  For instance, if $k_{\rm
gluon}$ is much the largest, then $p \to \Delta gg$, with $gg$ a spin-2
glueball, may be favored.  If on the other hand the right handed up
quark has the largest $k$, then $p \to \Delta^{++} \pi^-$ will probably
occur, and this possibility cannot be excluded on the basis of cosmic
rays, since the $\Delta^{++}$ is charged and would be the most stable
particle at large momenta.

\subsection{Dimension 6 operators}

If Lorentz symmetry is violated at a high scale, it is difficult to see
how to avoid a dispersion correction $E \sim p + O(p^3/\Lambda^2)$, with
$\Lambda$ a fundamental scale.  Presumably $\Lambda \leq m_{\rm pl}$ the
Planck scale, $m_{\rm pl} \simeq 1.2\times 10^{19}$ GeV.  For instance,
if spacetime is a lattice, then lattice dispersion relations of this
form are generically generated for free fields.%
\footnote
    {%
    Actually, if spacetime is a lattice, then interactions will
    generically induce $O(\alpha)$ corrections in dimension 4 Lorentz
    violating operators.  We have already seen that such corrections
    are ruled out by about 20 orders of magnitude.
    }

Simplemindedly, one might expect the most natural form for dimension 6
dispersion corrections to be a universal $K_1$ for all gauge fields and
a $K_{1/2}$ for all fermions.  Let us see what limits can be set in this
case.  If they are of comparable size, then a negative $K_1$ leads to
photon production, $p \to p \gamma$, while a negative $K_{1/2}$ leads to
$p \to p e^+ e^-$ (or neutrinos).  Table \ref{Table:dim6} shows that
only $10\%$ of a photon's $p^3$ weighted content is fermionic, while for
the electron less than $5\%$ is gauge boson; while both particles have
over 100 times the size of dimension 6 correction that hadrons have.  We
therefore neglect the ``cross-contamination'' of $\gamma$ and $e$, and
drop the small dimension 6 content of the hadrons.  For the process $p
\to p \gamma$, the process occurs when the energy shift,
\st
\Delta E = \frac{m_p^2}{2p}\frac{-x}{1-x} + 0.7 K x^3 p^3 \, ,
\stp
is positive.  Here 0.7 is $\int x^3 \gamma(x) dx$ for a photon, from
Table \ref{Table:dim6}.  The resulting bound is,
$K_1 >-1/(4\times 10^{22}\GeV)^{-2}$.
For the case $p \to p e^+ e^-$, we may neglect the
electron mass and consider the case where one electron emerges with very
small $x$; approximately the same limit is obtained,
$K_{1/2} > -1/(4\times 10^{22}\GeV)^{-2}$.
The scale appearing here is more than 2 orders of magnitude larger than
the Planck scale.

For the opposite case, $K_1$ or $K_{1/2}$ larger than 0, the proton
energy rises faster than linear with $p$ at large momenta, so ordinary
Cherenkov radiation of soft photons is allowed, $p \to p \gamma$, above
a threshold of
\st
.0008K_1 + .0047K_{1/2} = \frac{m^2}{6 p^4} \, ,
\stp
where the numerical constants are the gauge boson and fermion dimension 6
content of the proton from Table \ref{Table:dim6}.  The numerical values
of the bounds are,
$K_1 < (6\times 10^{21}\GeV )^{-2}$,
$K_{1/2} < (1.6\times 10^{22}\GeV)^{-2}$.  Surprisingly, these
bounds are only about 4 times weaker; this is due to the more favorable
kinematics for the $p \to p \gamma$ reaction in the positive $K$
scenario.

What if the dimension 6 operators show more species dependence?  The
cases $K_\gamma < 0$, $K_{\rm lepton}<0$ have the same 
constraints just found; so do the cases $K_g>0$, $K_q>0$.
For the case $K_q < 0$ but still universal between flavors, the dominant
reaction is $p \to p D^0$, which is allowed by virtue of the larger $x$
quark content of the $D^0$ meson.  The bound, derived in perfect analogy
with the bounds found so far, is
$K_q >-(6\times 10^{21}\GeV)^{-2}$.  If $K$ only applies to first
generation quarks, the strongest bound is from $p \to p \pi^0$, and the
bound is about 3 times weaker (but still exists).  A similar bound on
$K_g$ arises from $p \to p D^0$, $K_g > -(2\times 10^{21}\GeV)^{-2}$.
If we assume $K_q,K_g=0$, we can also get bounds on
positive $K_\gamma$, $K_e$ from $p \to p \gamma$ (with $\gamma$ soft),
because of the photon and electron content of
the proton; these bounds are
$K_\gamma < (3\times 10^{21}\GeV)^{-2}$,
$K_e < (1\times 10^{21}\GeV)^{-2}$.  If $K_g$, $K_q$ are nonzero,
then the bounds are,
$K_\gamma < 25 |K_q|$, $K_e < 500|K_q|$ (or the above bound, whichever
is weaker).  Even these much weaker bounds suggest a trans-Planckian
scale.

\section{Conclusions}

We have shown that the mere existence of hadronic, high energy cosmic
rays places constraints on Lorentz violating extensions of
the Standard Model which are tighter than any previous constraints for a
number of operators.  To find these constraints, we first argued that
Lorentz violating operators modify particle dispersion relations
in a way determined by the partonic content of the particle in
question.  Then, we determined the partonic content of various hadrons,
the photon, and the electron at the relevant scale, set by the energy of
the cosmic rays in question.  We then looked for Cherenkov-like
processes made kinematically possible by these dispersion corrections,
and excluded Lorentz violating processes which would prevent extremely
high energy, hadronic cosmic rays from existing.

We find that, generally, species dependence in the maximal propagation
speed of a particle, induced by dimension 4 Lorentz violating operators,
is constrained to be smaller than
a part in $10^{21}$.  This applied for surprisingly many such dispersion
corrections, and all of our limits are two-sided.  Limits on dimension
5, CPT violating operators imply that, if generated with $O(1)$
coefficients at the scale of Lorentz symmetry breaking,
their intrinsic scale must exceed
$10^{34}$ GeV;
for dimension 6 operators, the intrinsic scale must
exceed $10^{21}$ GeV.  It is easy to evade the bounds on dimension 5
operators; they are absent in theories with CPT invariance.  The limits
on propagation speed differences and dimension 6 operators are harder to
avoid in a theory with maximal Lorentz violation in the UV.
(Similarly, maximum speed differences in theories without Lorentz
symmetry are expected to be
$O(\alpha) \sim 10^{-2}$, as recently emphasized by Collins {\it et.\
al.\ } \cite{Collins}, but it might be possible that there is some
loophole in this argument.  The argument does not apply if Lorentz
symmetry is violated in some hidden sector, in which case the size of
Lorentz violating operators in the Standard Model could be much smaller.)

We should finish by commenting on some loopholes and limitations of our
analysis.  First, the loopholes.  Most but not all combinations of Lorentz
violating operators can be excluded.  In particular, we find the
interesting possibility that, if right-handed up quarks have a slower
propagation speed than any other particle (either due to dimension 4
corrections or higher order in momentum dependent corrections), then the
$\Delta^{++}$ particle can be absolutely stable at high energies.  This
possibility may also evade the GZK cutoff.
This is an example of the narrow, special corners of
the parameter space of Lorentz violating effects which are consistent
with observed high energy cosmic ray events.

Also note that not all forms of Lorentz
violation lead to dispersion corrections of the form that we have
bounded; for instance, spacetime noncommutativity can be understood in
terms of the introduction of high dimension, Lorentz violating operators
\cite{Carroll}, but at leading order, the operators in question have no
twist-two content (see \cite{Carroll}),
so they do not modify dispersion relations.  We therefore do not obtain
strong constraints on spacetime noncommutativity from our results
(although this will probably change if we work
beyond leading order in the parameter describing the noncommutativity).

One major limitation of our work is that we have considered only the
case where there is a single preferred direction; Kosteleck\'y's
Standard Model Extension \cite{Kostelecky} allows more general
corrections.  Second, we have not been very exhaustive about what sorts
of Lorentz violating operators to allow, for instance, generally taking
the corrections of all quark flavors to be the same.  Third, we have
worked within the Standard Model, neglecting the possibility, for
instance, that supersymmetry is realized at an energy scale not far
above the electroweak one.  This latter possibility would require the
inclusion of several additional parton distribution functions in our
evolution; we would find that the proton and other hadrons have
nontrivial gluino and squark content.  The tendency of hadronic PDFs to
resemble each other would also be strengthened, because $\alphas$ would
run more slowly and the set of splitting processes available would be
larger; therefore limits on, for instance, the quark-gluon velocity
difference would be weakened.  However, in general we do not believe
that this case would be very different from the case of the Standard
Model, because we expect the Lorentz violating operators to respect
supersymmetry--otherwise, how would we obtain low energy supersymmetry
in the first place?  We leave these issues to future work.

\section*{Acknowledgements}

We would like to thank Michael Luke for useful conversations.
This work was supported in part by grants from the
Natural Sciences and Engineering Research Council of Canada and by le Fonds
Qu\'eb\'ecois de la Recherche sur la Nature et les Technologies.

\appendix
\section{Splitting functions and coefficients in the Standard Model}
\label{appendixA}

The general form of the DGLAP equations is,
\st
\frac{\mu^2 \partial f_i(x,\mu^2)}{\partial \mu^2} = 
	\sum_j \int_x^1 \frac{dy}{y} P_{i\leftarrow j}(y) f_j(x/y,\mu^2)
	\, .
\label{eq:DGLAP}
\stp
Here $f_i(x,q^2)$ are the parton distribution functions; $j$ is an index
over distinct species.  $P_{i \leftarrow j}(y)$ is the rate at which
species type $j$ splits to produce species $i$ with momentum fraction
$y$.  It is generally determined in a series expansion in the coupling,
beginning at $O(\alpha/4\pi)$.

The DGLAP splitting functions, at leading order, are determined by the
squared matrix element for a particle of momentum $p$ to go to a
particle of momentum $xp+p_\perp$ and a particle of
momentum $(1-x)p-p_\perp$, evaluated at leading order in $p_\perp^2$.
Namely, if the matrix element (using the on-shell spinors or
perpendicular polarization states as if the particles were on shell)
squared, summed on final and averaged over initial spin states, is
$\M^2$, then the splitting function is given by
$\M^2 p^3 x(1-x)/ 2[ \p \times (x\p{+}\p_\perp) ]^2$.  That is, the
matrix element is to be divided by the squared invariant describing how
non-collinear the final states are, and multiplied by the energies of
the incoming and outgoing states (when using relativistic state
normalization).

For a Lagrangian containing a gauge field and $\nf$ Dirac fermions in
representation $f$ under this field,
\st
L = \frac{1}{4g^2} F_{\mu \nu} F^{\mu \nu} + \bar\psi (\nott{D}+m)
	\psi \, ,
\stp
we evaluate the splitting function for the process $\psi \to \psi
\gamma$, with $\gamma$ carrying momentum fraction $x$, by evaluating
\st
\left|\vphantom{\Big(} \bar u[(1{-}x)p+p_\perp,\sigma_f] \nott{\epsilon}
	u[p,\sigma_i] \right|^2
\, ,
\stp
averaging over initial spins and colors and summing over final spins and
colors.  Here $u[p,\sigma_i]$ is the spinor for a particle of momentum
$p$ and spin $\sigma$.  A straightforward and well known calculation
leads to the splitting functions, 
\bg
P_{\psi \leftarrow \psi}(y) & = & 2\cf
	\left( \frac{2}{(1-y)_+} - 1 - y + \frac{3}{2}\delta(1-y) \right)
	\, , \nn \\
P_{\psi \leftarrow \gamma}(y) & = & 4\tf \nf (1-2y+2 y^2) \, , \nn \\
P_{\gamma \leftarrow \psi}(y) & = & 2\cf \left( \frac{2}{y} - 2 + y \right)
	 \, , \nn \\
P_{\gamma \leftarrow \gamma}(y) & = & 4\ca \left( \frac{1}{(1-y)_+}
	+ \frac{1}{y} -2+y-y^2 \right) +
	\left( \frac{11}{3}\ca - \frac{4}{3} \tf \nf \right) \delta(1-y)
	\, ,
\nd
each times $\alpha/4\pi$.
Here $\cf$ is the quadratic Casimir of the fermionic representation
($\cf=4/3$ for SU(3), $3/4$ for SU(2), and $q^2$ for U(1)),
$\ca$ is the Casimir of the
adjoint representation ($\nc$ for SU($\nc$)), and $\tf$ is the fermion trace
normalization (1/2 for SU($\nc$), $q^2$ for U(1)).  Contrary to common
usage, the PDF for the fermion is taken in the above to be the sum of the usual
fermion and anti-fermion PDF.

If instead we consider a complex scalar field, the quantity to compute
is
\st
\Big[ ( p + xp+p_\perp ) \cdot \epsilon \Big]^2 \, ,
\stp
and splitting functions turn out to be $\alpha/4\pi$ times
\bg
P_{\phi \leftarrow \phi}(y) & = & 4\cs
	\left( \frac{1}{(1-y)_+} - 1 + \delta(1-y) \right)
	\, , \nn \\
P_{\phi \leftarrow \gamma}(y) & = & 8\ts \ns (y- y^2) \, , \nn \\
P_{\gamma \leftarrow \phi}(y) & = & 4\cs \left( \frac{1}{y} - 1 \right)
	 \, , \nn \\
P_{\gamma \leftarrow \gamma}(y) & = & \mbox{[above]}\;
	- \frac{2}{3} \ts \ns \: \delta(1-y) \, .
\nd

The DGLAP splitting functions arising from the Lagrangian terms,
\st
L = (D_\mu H)^\dagger (D^\mu H) + y_t
	\left( H \bar t_L P_R t_R + {\rm h.c.} \right) \, ,
\stp
arise from evaluating $|\bar u(xp+p_\perp) u(p)|^2$.  For the case where
all three fields, $H$, $t_R$, and $t_L$ are singlets, the splitting
functions are $y_t^2/16 \pi^2$ times
\bg
P_{\psi_i \leftarrow \psi_j}(y) & = & (1-y)(1-\delta_{ij})
	- \frac{1}{2} \delta(1-y) \delta_{ij} \, ,
\nn \\
P_{\psi_i \leftarrow \phi}(y) & = & 1 \, ,
\nn \\
P_{\phi \leftarrow \psi_i}(y) & = & y \, ,
\nn \\
P_{\phi \leftarrow \phi}(y) & = & - \delta(1-y) \, ,
\nd
where the indices $i,j$ refer to whether the field is $t_R$ or $t_L$.

The structure of \Eq{eq:DGLAP} means that the integral moments
of interest to us obey closed sets of ordinary differential RG
equations.  This point is ``well
known to the people who know it well,'' but may not be familiar to all
readers, so we go through it in a little detail.  Introduce
\st
f_{i,n}(q^2) \equiv \int_0^1 x^n f_i(x,q^2) dx \, ,
\stp
the moment of the PDF corresponding to the expectation value of a
dimension $n+3$, twist-2 operator.
We obtain an evolution equation for $f_{i,n}$ from \Eq{eq:DGLAP} by
integrating it over $x^n dx$;
\bg
\int_0^1 x^n \frac{q^2 df_{i}(x,q^2)}{dq^2} dx & = &
	\sum_j \int_0^1 x^n dx \int_x^1 \frac{dy}{y} P_{i \leftarrow j}(y)
	\;f_j(x/y,q^2) \nn \\
\frac{q^2 df_{i,n}(q^2)}{dq^2} & = &
	\sum_j \int_0^1 \frac{dy}{y} \int_0^y x^n dx \; P_{i \leftarrow j}(y)
	\;f_j(x/y,q^2) \nn \\
& = & \sum_j \int_0^1 y^n dy \int_0^1 z^n dz \; P_{i\leftarrow j}(y)
	\;f_j(z,q^2) \nn \\
& = & \sum_j \left( \int_0^1 z^n f_j(z,q^2) dz \right)
	\left( \int_0^1 y^n dy P_{i \leftarrow j}(y) dy \right) \nn \\
& = & \sum_j C_{ij,n} f_{j,n}(q^2) \, , \qquad \mbox{with} \quad
	C_{ij,n} \equiv \int_0^1 y^n P_{i\leftarrow j}(y) dy \, .
\nd
The evolution equations for the $f_{i,n}$ form a closed set of ordinary
differential renormalization group equations, which mix particle species
but not operator dimension $n$.

It should not be surprising
that the expectation values of all dimension $D$, twist 2 operators
should satisfy closed renormalization group equations in this way.  We
expect an operator to mix under renormalization group flow only with
operators of the same spin and equal or lower dimension.  Since twist 2
operators have the maximum spin available at a given dimension, the
twist 2 operators of dimension $D$ can only mix with other twist 2
operators at dimension $D$.

It remains to evaluate the $C_{ij}$ for all particles of the Standard
Model, from the splitting functions we have just given.  This is a
straightforward combinatorial problem.  We will
write the fermionic fields as $[E,L,U,D,Q]$ for the right-handed electron,
leptonic doublet, right-handed up, right-handed down, and quark doublet
fields, $[B,W,G]$ for the hypercharge, weak SU(2), and strong SU(3)
(gluon) gauge bosons, and $H$ for the Higgs doublet.  The couplings are
$[a',a,a_s,a_y]$, with $a=\alpha/4\pi$ and $a_y = y^2/16\pi^2$ with $y$
the Yukawa coupling of the top quark, $m_t^2 = y^2 v^2/2$ ($v$ the Higgs
vacuum expectation value, $G_F=1/v^2 \sqrt{2}$).  The evolution
equations for the integral moments of the PDFs are, at leading order,
\bg
\frac{q^2 d}{dq^2} E_m & = & a' C_{ff} E_m + a' C_{fg} B
	\, , \nn \\
\frac{q^2 d}{dq^2} L_m & = & \left( \frac{1}{4} a' + \frac{3}{4} a \right)
	C_{ff} L_m + \frac{1}{2} a C_{fg} W + \frac{1}{2} a' C_{fg} B
	\, , \nn \\
\frac{q^2 d}{dq^2} D_m & = & \left( \frac{4}{3} a_s + \frac{1}{9} a'
	\right) C_{ff} D_m + \frac{1}{2} C_{fg} a_s G + \frac{1}{3}
	C_{fg} a' B
	\, , \nn \\
\frac{q^2 d}{dq^2} U_m & = & \left( \frac{4}{3} a_s + \frac{4}{9} a'
	\right) C_{ff} U_m + \frac{1}{2} C_{fg} a_s G + \frac{4}{3}
	C_{fg} a' B
	\nn \\ &&
	+ a_y \delta_{m3} \left[ 2 C_{ffy} U_3 + C_{f2fy} Q_3 + C_{fs} H
	\right]
	\, , \nn \\
\frac{q^2 d}{dq^2} Q_m & = & \left( \frac{4}{3} a_s + \frac{3}{4} a
	+ \frac{1}{36} a' \right) C_{ff} Q_m + \frac{1}{2} C_{fg} a_s G
	+ \frac{3}{2} C_{fg} a W + \frac{1}{6} C_{fg} a' B
	\nn \\ &&
	+ a_y \delta_{m3} \left[ C_{ffy} Q_3 + 2C_{f2fy} U_3 + C_{fs} H
	\right]
	\, , \nn \\
\frac{q^2 d}{dq^2} B & = & a' C_{gf} \sum_m \left( E_m + \frac{1}{4} L_m
	+ \frac{1}{9} D_m + \frac{4}{9} U_m + \frac{1}{36} Q_m \right)
	+ \frac{10 \ng}{3} a' C_{ggf} B
	\nn \\ &&
	+ \frac{1}{4} a' C_{gs} H + \frac{1}{2} a' C_{ggs} B
	\, , \nn \\
\frac{q^2 d}{dq^2} W & = & a C_{gf} \sum_m \left( \frac{3}{4} L_m
	+ \frac{3}{4} Q_m \right) + 2\ng a C_{ggf} W + 2 a C_{ggg} W
	\nn \\ &&
	+ \frac{3}{4} a C_{gs} H + \frac{1}{2} a C_{ggs} W
	\, , \nn \\
\frac{q^2 d}{dq^2} G & = & a_s C_{gf} \sum_m \left( \frac{4}{3} U_m
	+ \frac{4}{3} D_m + \frac{4}{3} Q_m \right)
	+ 2 \ng a_s C_{ggf} G + 3 a_s C_{ggg} G
	\, , \nn \\
\frac{q^2 d}{dq^2} H & = & \left( \frac{3}{4} a + \frac{1}{4} a'
	\right) C_{ssg} H + \frac{1}{2} a C_{sg} W
	+ \frac{1}{2} a' C_{sg} B
	\nn \\ &&
	+ a_y \left[ C_{ssy} H + C_{sf} ( Q_3 + 2 U_3 ) \right]
	\, ,
\nd
where $m$ is a generation index, $N_g=3$ is the number of generations,
and the $C_{xxx}$ coefficients are integrals of splitting functions
given explicitly by,
\bg
&&C_{ff} = \frac{-8}{3} \, , \quad
C_{gf} = \frac{8}{3} \, , \quad
C_{fg} = \frac{2}{3} \, , \quad
C_{ggf} = \frac{-2}{3} \, , \quad
C_{ggg} = 0 \, ,
	\nn \\
&&C_{ssg} = -2 \, , \quad
C_{gs} = 2 \, , \quad
C_{sg} = \frac{1}{3} \, , \quad
C_{ggs} = \frac{-1}{3} \, ,
	\nn \\
&&C_{ffy} = \frac{-1}{2} \, , \quad
C_{f2fy} = \frac{1}{6} \, , \quad
C_{sf} = \frac{1}{3} \, , \quad
C_{fs} = \frac{1}{2} \, , \quad
C_{ssy} = -1 \, ,
\nd
for the case of the dimension 4 integral moments and
\bg
&&C_{ff} = \frac{-157}{30} \, , \quad
C_{gf} = \frac{11}{15} \, , \quad
C_{fg} = \frac{11}{30} \, , \quad
C_{ggf} = \frac{-2}{3} \, , \quad
C_{ggg} = \frac{-21}{5} \, ,
	\nn \\
&&C_{ssg} = \frac{-13}{3} \, , \quad
C_{gs} = \frac{1}{3} \, , \quad
C_{sg} = \frac{2}{15} \, , \quad
C_{ggs} = \frac{-1}{3} \, ,
	\nn \\
&&C_{ffy} = \frac{-1}{2} \, , \quad
C_{f2fy} = \frac{1}{20} \, , \quad
C_{sf} = \frac{1}{5} \, , \quad
C_{fs} = \frac{1}{4} \, , \quad
C_{ssy} = -1 \, ,
\nd
for the case of dimension 6 integral moments.  The $a_s^2$ corrections
can be found in \cite{Vermaseren}, and the Standard Model beta functions
are in \cite{Ford}, and we will not repeat them here.

\end{document}